\documentclass{article}
\usepackage[latin9]{inputenc}
\usepackage{amsmath}

\makeatletter

\newcommand{\noun}[1]{\textsc{#1}}
\DeclareTextSymbolDefault{\textquotedbl}{T1}

\newcommand{\lyxaddress}[1]{
	\par {\raggedright #1
	\vspace{1.4em}
	\noindent\par}
}

\makeatother

\begin{document}
\title{\textbf{On Bethe Ansatz for a Supersymmetric Vertex Model with $\mathcal{U_{\textrm{q}}}$$[osp(2|2)^{(2)}]$
symmetry}}
\author{G. K. Sampa and A. Lima-Santos}
\maketitle

\lyxaddress{Universidade Federal de São Carlos, Departamento de Física, Caixa
Postal 676, CEP 13569-905, São Carlos, Brasil}
\begin{abstract}
The Algebraic Bethe ansatz for a supersymmetric nineteen vertex-model
constructed from a three-dimensional representation of the twisted
quantum affine Lie superalgebra $\mathcal{U}_{q}[\mathrm{osp}(2|2)^{(2)}]$
is presented in detail. The eigenvalues and eigenvectors of the row-to-row
transfer matrix are calculated and the corresponding Bethe Ansatz
equations are obtained and analyzed numerically. 
\end{abstract}
\vspace{1.2cm}
 
\begin{center}
\par\end{center}

\newpage{}

\section{Introduction}

One-dimensional quantum spin chain Hamiltonians and classical statistical
systems in two spatial dimensions on a lattice (vertex models), share
a common mathematical structure responsible by our understanding of
these integrable models \cite{Baxter,KIB}. If the Boltzmann weights
underlying the vertex models are obtained from solutions of the Yang-Baxter
({\small{}YB}) equation the commutativity of the associated transfer
matrices immediately follow, leading to their integrability.

The Bethe Ansatz ({\small{}BA}) is the powerful method in the analysis
of integrable quantum models. There are several versions: Coordinate
{\small{}BA \cite{Bethe}} ,Algebraic{\small{}\ BA \cite{FT} },
Analytical {\small{}BA \cite{VR} }, etc. developed for diagonalization
of the corresponding Hamiltonian.

The simplest version is the Coordinate {\small{}BA }which we can obtain
the eigenfunctions and the spectrum of the Hamiltonian from its eigenvalue
problem. It is really simple and clear for the two-state models like
the six-vertex models but becomes awkward for models with a higher
number of states.

The Algebraic {\small{}BA}, also proverbial as Quantum Inverse Scattering
method, is an elegant and important generalization of the Coordinate
{\small{}BA}. It is based on the idea of constructing eigenfunctions
of the Hamiltonian via creation and annihilation operators acting
on a reference state. Here we use the fact the {\small{}YB} equation
can be recast in the form of commutation relations for the matrix
elements of the monodromy matrix which play the role of creation and
annihilation operators. From this monodromy matrix we get the transfer
matrix which, by construction, commutes with the Hamiltonian. Thus,
constructing eigenfunctions of the transfer matrix determines the
eigenfunctions of the Hamiltonian.

Imposing appropriate boundary conditions the {\small{}BA} method leads
to a system of equations, the {\small{}BA} equations, which are useful
in the thermodynamic limit. The energy of the ground state and its
excitations, velocity of sound, etc., may be calculated in this limit.
Moreover, in recent years we witnessed another very fruitful connection
between the {\small{}BA} method and conformal field theory. \ Using
the Algebraic {\small{}BA}, Korepin \cite{KO} found various representations
of correlators in integrable models. Moreover Babujian and Flume \cite{BF}
developed a method from the Algebraic {\small{}BA} which reveals a
link to the Gaudin model and render in the quasiclassical limit solutions
of the Knizhnik-Zamolodchikov equations for the $SU(2)$ Wess-Zumino-Novikov-Witten
conformal theory.

Integrable quantum systems containing Fermi fields have been attracting
increasing interest due to their potential applications in condensed
matter physics. The prototypical examples of such systems are the
supersymmetric generalizations of the Hubbard and $t$-$J$ models
\cite{EK0}, which play an important role in condensed matter physics,
and also the search for solutions of the graded \emph{Yang-Baxter
equatio}ns {[}{]} which gave origin to important algebraic construction
as the supersymmetric Hopf algebras and quantum groups \cite{drinf}.
More recently, the integrability of supersymmetric models also proved
to be important in superstring theory, more specifically in the AdS/CFT
correspondence \cite{mald,beisert,bena}. They lead to a generalization
of the {\small{}YB} equation associated with the introduction of the
a $Z_{2}$- grading \cite{KS1} in the {\small{}YB} equation.

In the context of the Algebraic {\small{}BA} , the version presented
here is based on the Tarosov approach \cite{Ta}

The paper is organized as follows: In Section $2$ we present the
models. In Section $3$ he main steps of the algebraic {\small{}BA
are developed in detail in order to solve the eigenvalue problem for
the row to row transfer matrix with periodic boundary conditions,
where its eigenvectors and eigenvalues are presented to be fixed by
the roots of the Bethe equations. In section 4 we made a numecal analysis
with the Bethe equation, and Section 5 is for our closing remarks.}{\small\par}

\section{The model}

The most powerful and beautiful method to analyze these integrable
quantum systems probably is the \emph{Algebraic}\textsc{ }{\small{}BA}
\cite{FT}. This technique allows one to diagonalize the transfer
matrix of a given integrable quantum system in an analytical way.
The \textsc{aba} was originally applied to systems with periodic boundary
conditions but after the work of Sklyanin \cite{skly}, integrable
models with non-periodic boundaries could also be handled.

In this work we will study another graded three-state model with periodic
boundary conditions. The $R$-matrix associated with this model is
constructed from a three-dimensional free boson representation $V$
of the twisted quantum affine Lie superalgebra $U_{q}[\mathrm{osp}(2|2)^{(2)}]\simeq U_{q}[C(2)^{(2)}]$
. We would like to emphasize that vertex-models described by Lie superalgebras
$-$ and, in particular, by twisted Lie superalgebras $-$ are usually
the most complex ones, which is due, of course, to the high complexity
of such Lie superalgebras \cite{frap1,frap2,khoro,mack,rans,xuzhang}.

Let $\text{\text{W=V\ensuremath{\oplus}U}}$ be a $Z_{2}$-graded
vector space where $V$ and $U$ denote its even and odd parts, respectively.
In a $Z_{2}$-graded vector space we associate a gradation $p(i)$
to each element $\epsilon_{i}$ of a given basis of $V$. In the present
case, we shall consider only a three-dimensional representation of
the twisted quantum affine Lie superalgebra $U_{q}[\mathrm{osp}(2|2)^{(2)}]$
with a basis $E=\{\epsilon_{1},\epsilon_{2},\epsilon_{3}\}$ and the
grading $p(1)=0$, $p(2)=1$ and $p(3)=0$. Multiplication rules in
the graded vector space $W$ differ from the ordinary ones by the
appearance of additional signs. For example, the graded tensor product
of two homogeneous even elements $A\in\mathrm{End}(V)$ and $B\in\mathrm{End}(V)$
turns out to be defined by the formula, 
\begin{equation}
A\otimes^{g}B=\sum_{i,j,k,l=1}^{d}\left(-1\right)^{p(i)p(k)+p(j)p(k)}A_{ij}B_{kl}\left(e_{ij}\otimes e_{kl}\right),\label{int.1}
\end{equation}
where $d$ (in the present case, $d=3$) is the dimension of the vector
space $V$ and $e_{ij}$ are the Weyl matrices ($e_{ij}$ is a matrix
in which all elements are null, except that element on the $[i,j]$
position, which equals $1$). In the same fashion, the graded permutation
operator $P^{g}$ is defined by 
\begin{equation}
P^{g}=\sum_{i,j=1}^{3}\left(-1\right)^{p(i)p(j)}\left(e_{ij}\otimes e_{ji}\right).\label{int.2}
\end{equation}
and the graded transposition $A^{t^{g}}$ of a matrix $A\in\mathrm{End}(V)$
as well as its inverse graded transposition, $A^{\tau^{g}}$, are
defined, respectively, by 
\begin{equation}
A^{t^{g}}=\sum_{i,j=1}^{3}\left(-1\right)^{p(i)p(j)+p(i)}A_{ji}e_{ij},\qquad A^{\tau^{g}}=\sum_{i,j=1}^{3}\left(-1\right)^{p(i)p(j)+p(j)}A_{ji}e_{ij},\label{int.3}
\end{equation}
so that $A^{t^{g}\tau^{g}}=A^{\tau^{g}t^{g}}=A$. Finally, the graded
trace of a matrix $A\in\mathrm{End}(V)$ is given by 
\begin{equation}
\mathrm{tr}^{g}(A)=\sum_{i=1}^{3}\left(-1\right)^{p(i)}A_{ii}e_{ii}.
\end{equation}

The YB equation {[}{]},

\begin{equation}
\mathcal{R}_{12}(x)\mathcal{R}_{13}(xy)\mathcal{R}_{23}(y)=\mathcal{R}_{23}(y)\mathcal{R}_{13}(xy)\mathcal{R}_{12}(x),\label{YBE}
\end{equation}
is written in the same way as in in the non-graded case: it is only
necessary to employ graded operations instead of the usual operations 

The $R$-matrix, solution of the graded YB equation (\ref{YBE}),
associated with the Yang-Zhang vertex-model \cite{YZ} can be written,
up to a normalizing factor and employing a different notation, as
follows: 
\begin{equation}
\mathcal{R}(x)=\left(\begin{array}{ccccccccc}
r_{1}(x) & 0 & 0 & 0 & 0 & 0 & 0 & 0 & 0\\
0 & r_{2}(x) & 0 & r_{5}(x) & 0 & 0 & 0 & 0 & 0\\
0 & 0 & r_{3}(x) & 0 & r_{6}(x) & 0 & r_{7}(x) & 0 & 0\\
0 & s_{5}(x) & 0 & r_{2}(x) & 0 & 0 & 0 & 0 & 0\\
0 & 0 & s_{6}(x) & 0 & r_{4}(x) & 0 & r_{6}(x) & 0 & 0\\
0 & 0 & 0 & 0 & 0 & r_{2}(x) & 0 & r_{5}(x) & 0\\
0 & 0 & s_{7}(x) & 0 & s_{6}(x) & 0 & r_{3}(x) & 0 & 0\\
0 & 0 & 0 & 0 & 0 & s_{5}(x) & 0 & r_{2}(x) & 0\\
0 & 0 & 0 & 0 & 0 & 0 & 0 & 0 & r_{1}(x)
\end{array}\right),\label{R}
\end{equation}
where the amplitudes $r_{i}\left(x\right)$ and $s_{i}\left(x\right)$
are given respectively by 
\begin{eqnarray}
r_{1}\left(x\right) & = & q^{2}x-1,\\
r_{2}\left(x\right) & = & q\left(x-1\right),\\
r_{3}\left(x\right) & = & q\left(q+x\right)\left(x-1\right)/\left(qx+1\right),\\
r_{4}\left(x\right) & = & q\left(x-1\right)-\left(q+1\right)\left(q^{2}-1\right)x/\left(qx+1\right),\\
r_{5}\left(x\right) & = & q^{2}-1,\\
r_{6}\left(x\right) & = & -q^{1/2}\left(q^{2}-1\right)\left(x-1\right)/\left(qx+1\right),\\
r_{7}\left(x\right) & = & \left(q-1\right)\left(q+1\right)^{2}/\left(qx+1\right),\\
s_{5}\left(x\right) & = & \left(q^{2}-1\right)x=xr_{5}\left(x\right),\\
s_{6}\left(x\right) & = & -q^{1/2}\left(q^{2}-1\right)x\left(x-1\right)/\left(qx+1\right)=xr_{6}\left(x\right),\\
s_{7}\left(x\right) & = & \left(q-1\right)\left(q+1\right)^{2}x^{2}/\left(qx+1\right)=x^{2}r_{7}\left(x\right).
\end{eqnarray}

This $R$-matrix has the following properties or symmetries {[}27{]}:
\begin{eqnarray}
 & \text{regularity:}\qquad & \mathcal{R}_{12}\left(1\right)=f\left(1\right)^{1/2}P_{12}^{g},\label{Sym1}\\
 & \text{unitarity:} & \mathcal{R}_{12}\left(x\right)=f\left(x\right)\mathcal{R}_{21}^{-1}\left(x^{-1}\right),\label{Sym2}\\
 & \text{super PT:} & \mathcal{R}_{12}\left(x\right)=\mathcal{R}_{21}^{t_{1}^{g}\tau_{2}^{g}}\left(x\right),\label{Sym3}\\
 & \text{crossing:} & \mathcal{R}_{12}\left(x\right)=g\left(x\right)\left[V_{1}\mathcal{R}_{12}^{t_{2}^{g}}\left(\eta^{-1}x^{-1}\right)V_{1}^{-1}\right],\label{Sym4}
\end{eqnarray}
where 
\begin{equation}
f\left(x\right)=r_{1}(x)r_{1}\left(\frac{1}{x}\right)=\left(q^{2}x-1\right)\left(\frac{q^{2}}{x}-1\right),\qquad g(x)=-\frac{qx\left(x-1\right)}{\left(qx+1\right)}.
\end{equation}
Here, $t_{1}^{g}$ and $t_{2}^{2}$ mean graded partial transpositions
in the first and second vector spaces, respectively; $\tau_{1}^{2}$
and $\tau_{2}^{2}$ the corresponding inverse operations. Besides,
$\eta=-q$ is the \emph{crossing parameter} while 
\begin{equation}
M=V^{t^{g}}V=\mathrm{diag}\left(1/q,1,q\right)
\end{equation}
is the \emph{crossing matrix}.

Besides ${\cal R}$ we have to consider matrices $R=P^{g}\mathcal{R}$
which satisfy 
\begin{equation}
R_{12}(x)R_{23}(xy)R_{12}(y)=R_{23}(y)R_{12}(xy)R_{23}(x)\label{eq2.5}
\end{equation}
Because only $R_{12}$ and $R_{23}$ are involved, Eq.(\ref{eq2.5})
written in components looks the same as in the non graded case.

\section{The Algebraic Bethe Ansatz}

In the previous section we have presented the model through its $\mathcal{\mathcal{\textrm{R}}}$-matrix 

The main problem now is the diagonalization of the transfer matrix
of the lattice system. To do this we recall the formulation of the
Algebraic Bethe ansatz \cite{Ta}.

Let us consider a regular lattice with $L$ columns and $L^{\prime}$
rows. A physical state on this lattice is defined by the assignment
of a {\em state variable }to each lattice edge. If one takes the
horizontal direction as space and the vertical one as time, the transfer
matrix plays the role of a discrete evolution operator acting on the
Hilbert space ${\cal H}^{(N)}$ spanned by the {\em row states}
which are defined by the set of vertical link variables on the same
row. Thus, the transfer matrix elements can be understood as the transition
probability of the one row state to project on \ the consecutive
one after a unit of time.

The standard row-to-row monodromy matrix for an L- tensor space
\begin{equation}
V^{(1)}\otimes V^{(2)}\otimes\cdots\otimes V^{(L)}
\end{equation}
\begin{equation}
T_{0}(x)=\mathcal{R}_{0L}(x)\mathcal{R}_{0N-1}(x)\cdots\mathcal{R}_{01}(x)
\end{equation}

A quantum integrable system is characterized by monodromy matrix $T_{0}(x)$
satisfying the equation 
\begin{equation}
R_{00'}(x/w)\left[T_{0}(x)\stackrel{g}{\otimes}T_{0'}(w)\right]=\left[T_{0'}(w)\stackrel{g}{\otimes}T_{0}(x)\right]R_{00'}(x/w)\label{eq4.1}
\end{equation}
whose consistency is guaranteed by the {\small{}YB} equation (\ref{eq2.5}).
$T_{0}(x)$ is a matrix in the quantum space $\otimes_{j}^{N}$$V^{(j)}$
with matrix elements that are operators on the states of the quantum
system. The space $V_{(0)}$ is called auxiliary space of $T_{0}(x)$.

From the auxiliary space we can see $T_{0}(x$) as a matrix $3$ by
$3$ 
\begin{equation}
T_{0}(x)=\left(\begin{array}{ccc}
A_{1}(x) & B_{1}(x) & B_{2}(x)\\
C1(x) & A_{2}(x) & B_{3}(x)\\
C_{2}(x) & C_{3}(x) & A_{3}(x)
\end{array}\right)
\end{equation}
where the operators $A_{i},B_{i},C_{i}$ are $3^{L}$by $3^{L}$ matrices.

The transfer matrix$\tau(x)$ for periodic boundary condition is defined
as the super-trace of the row-to-row monodromy

\begin{equation}
\tau(x)={\rm Str}T_{0}(x)=\sum_{i=1}^{3}(-1)^{[i]}\ A_{i}(x)=A_{1}(x)-A_{2}(x)+A_{3}(x)\label{eq4.6}
\end{equation}
In particular, the Hamiltonians can also be derived by the well-known
relation 
\begin{equation}
H=\alpha\ \frac{\partial}{\partial x}\left(\ln\tau(x)\right)_{x=1}\label{eq4.7}
\end{equation}

In this section we will derive the {\small{}BA} equations of $19$-vertex
models presented in Section $2$ using the Algebraic {\small{}BA}
developed by Tarasov \cite{Ta}. To do this we need of the commutation
relations for entries of the monodromy matrix which are derived from
the fundamental relation (\ref{eq4.1}). Here these commutation relations
do not share a common structure. Therefore, we only write some of
them in the text and recall (\ref{eq4.1}) to get the remaining ones.

First of all , let us observe that for each row state one can define
the magnon number operator which commutes with the transfer matrix
of the models 
\begin{equation}
[\tau(x),M]=0,\quad M=\sum_{k=1}^{L}M_{k},\qquad M_{k}=\left(\begin{array}{lll}
0 & 0 & 0\\
0 & 1 & 0\\
0 & 0 & 2
\end{array}\right),\label{eq4.8}
\end{equation}
This is the analog of the operator $S_{T}^{z}$ used in the previous
section and the relation between $M$ and the spin total $S_{T}^{z}$
is simply $M=L-S_{T}^{z}$. Once again, the Hilbert space can be broken
down into sectors ${\cal H}_{M}^{(L)}$. In each of these sectors,
the transfer matrix can be diagonalized independently, $\tau(x)\Psi_{M}=\Lambda_{M}(\{x_{i}\}\Psi_{M}$,$\:(i=0,1,...,M),$$x_{0}=x$.
We now start to diagonalize $\tau(x)$ in every sector:

\subsection{Sector $M=0$}

Let us consider the highest vector of the monodromy matrix $T(u)$
in a lattice of $L$ sites as the even (bosonic) completely unoccupied
state 
\begin{equation}
\Psi_{0}\equiv\left|0\right\rangle =\otimes_{k=1}^{L}\left(\begin{array}{l}
1\\
0\\
0
\end{array}\right)_{k}\label{eq4.9}
\end{equation}
It is the only state in the sector with $M=0$. Using (27) we can
compute the action of the matrix elements of $\tau(x)$ on this reference
state: 
\begin{eqnarray}
\quad A_{k}(x)\left|0\right\rangle  & = & r_{k}^{L}(x)\left|0\right\rangle ,\nonumber \\
\quad C_{k}(x)\left|0\right\rangle  & = & 0,\quad B_{k}(x)\left|0\right\rangle \neq\{0,\left|0\right\rangle \},\quad k=1,2,3\label{eq4.10}
\end{eqnarray}
Therefore in the sector $M=0$ , $\Psi_{0}$ is the eigenstate of
$\tau(x)=A_{1}(x)-A_{2}(x)+A_{3}(x)$ with eigenvalue 
\begin{equation}
\Lambda_{0}(x)=r_{1}^{L}(x)-r_{2}^{L}(x)+r_{3}^{L}(x)\label{eq4.11}
\end{equation}
Here we observe that the action of the operators $B_{i}(x)$ on the
reference state will give us new states which lie in sectors with
$M\neq0$.

\subsection{Sector $M=1$}

In this sector we have the states $B_{1}\left|0\right\rangle $ and
$B_{3}\left|0\right\rangle $. Since $B_{3}\left|0\right\rangle \propto B_{1}\left|0\right\rangle $,
we seek eigenstate of the form 
\begin{equation}
\Psi_{1}(x_{1})=B_{1}(x_{1})\left|0\right\rangle .\label{eq4.12}
\end{equation}

The action of the operator $\tau(x)$ on this state can be computed
with aid of the following commutation relations 
\begin{eqnarray}
A_{1}(x)B_{1}(w) & = & z(w/x)B_{1}(w)A_{1}(x)-\frac{r_{5}(w/x)}{r_{2}(w/x)}B_{1}(x)A_{1}(w)\label{eq4.13a}\\
A_{2}(x)B_{1}(w) & = & -\frac{z(x/w)}{\omega(x/w)}B_{1}(w)A_{2}(x)-\frac{z(x/w)}{\omega(x/w)}\frac{1}{y(w/x)}B_{2}(w)C_{1}(x)\nonumber \\
 &  & +\frac{s_{5}(x/w)}{r_{2}(x/w)}B_{1}(x)A_{2}(w)+\frac{s_{5}(x/w)}{r_{2}(x/w)}\frac{1}{y(x/w)}B_{2}(x)C_{1}(w)\nonumber \\
 &  & +\frac{1}{y(x/w)}B_{3}(x)A_{1}(w)\\
A_{3}(x)B_{1}(w) & = & \frac{r_{2}(x/w)}{r_{3}(x/w)}B_{1}(w)A_{3}(x)+\frac{1}{y(x/w)}B_{3}(x)A_{2}(w)\nonumber \\
 &  & +\frac{r_{5}(x/w)}{r_{3}(x/w)}B_{2}(w)C_{3}(x)-\frac{s_{7}(x/w)}{r_{3}(x/w)}B_{2}(x)C_{3}(w)\label{eq4.13}
\end{eqnarray}
where we have used Tarasov's notation\cite{Ta},for the ratio functions
\begin{eqnarray}
z(x) & = & \frac{r_{1}(x)}{r_{2}(x)},\quad\omega(x)=\frac{r_{1}(x)r_{3}(x)}{-r_{3}(x)r_{4}(x)+r_{6}(x)s_{6}(x)},\nonumber \\
\quad y(x) & = & \frac{r_{3}(x)}{s_{6}(x)}=,\quad y(x^{-1})=\frac{-r_{3}(x)r_{4}(x)+r_{6}(x)s_{6}(x)}{r_{7}(x)s_{6}(x)-r_{3}(x)r_{6}(x)},\label{eq4.14}
\end{eqnarray}

When $\tau(x)$ act on $\Psi_{1}(x_{1})$ , the corresponding eigenvalue
equation has two unwanted terms: 
\begin{eqnarray}
\tau(x)\Psi_{1}(x_{1}) & = & \left(A_{1}(x)-A_{2}(x)+A_{3}(x)\right)\Psi_{1}(x_{1})\nonumber \\
 & = & [z(x_{1}/x)r_{1}^{L}(x)+\frac{z(x/x_{1})}{\omega(x/x)}r_{2}^{L}(x)+\frac{r_{2}(x/x_{1})}{r_{3}(x/x_{1})}r_{3}^{L}(x)]\Psi_{1}(x_{1})\nonumber \\
 &  & -[\frac{r_{5}(x_{1}/x)}{r_{2}(x_{1}/x)}r_{1}^{L}(x_{1})+\frac{s_{5}(x/x_{1})}{r_{2}(x/x_{1})}r_{2}^{L}(x_{1})]B_{1}(x)\left|0\right\rangle \nonumber \\
 &  & -\frac{1}{y(x/x)}r_{1}^{L}(x_{1})-\frac{1}{y(x/x_{1})}r_{2}^{L}(x_{1})]B_{3}(x)\left|0\right\rangle \label{eq4.15}
\end{eqnarray}
From the matrix elements (7-16) we can see that $r_{5}(x)/r_{2}(x)=-s_{5}(x^{-1})/r_{2}(x^{-1})$.
Therefore the unwanted terms vanish and $\Psi_{1}(x_{1})$ is eigenstate
of $\tau(x)$ with eigenvalue 
\begin{equation}
\Lambda_{1}(x,x_{1})=z(x_{1}/x)r_{1}^{L}(x)+\frac{z(x/x_{1})}{\omega(x/x_{1})}r_{2}^{L}(x)+\frac{r_{2}(x/x_{1})}{r_{3}(x/x_{1})}r_{3}^{L}(x)\label{eq4.16}
\end{equation}
provided 
\begin{equation}
\left(z(x_{1})\right)^{L}=1\label{eq4.17}
\end{equation}

\subsection{Sector $M=2$}

In the sector $M=2$, we encounter two linearly independent states
$B_{1}B_{1}\left|0\right\rangle $ and $B_{2}\left|0\right\rangle $.
(The states $B_{3}B_{3}\left|0\right\rangle ,B_{1}B_{3}\left|0\right\rangle $
and $B_{3}B_{1}\left|0\right\rangle $ also lie in the sector $M=2$
but they are proportional to the state $B_{1}B_{1}\left|0\right\rangle $).
We seek eigenstates in the form 
\begin{equation}
\Psi_{2}(x_{1},x_{2})=B_{1}(x_{1})B_{1}(x_{2})\left|0\right\rangle +B_{2}(x_{1})\Gamma(x_{1},x_{2})\left|0\right\rangle \label{eq4.18}
\end{equation}
where $\Gamma(x_{1},x_{2})$ is an operator-valued function which
has to be fixed such that $\Psi_{2}(x_{1},x_{2})$ is unique state
in the sector $M=2$.

Here we observe that the operator-valued function $\Gamma(x_{1},x_{2})$
is.

It was demonstrated in \cite{TA} that $\Psi_{2}(x_{1},x_{2})$ is
unique provided it is ordered in a normal way: In general, the operator-valued
function $\Psi_{n}(x_{1},...,x_{n})$ is composite of normal ordered
monomials. A monomial is normally ordered if in it all elements of
the type $B_{i}(x)$ are on the left, and all elements of the type
$C_{j}(x)$ on the right of all elements of the type $A_{k}(x).$
Moreover, the elements of one given type having standard ordering:
$T_{i_{1}j_{1}}(x_{1})T_{i_{2}j_{2}}(x_{2})...T_{i_{n}jn}(x_{n})$.
For a given sector $M=$ $n$, $\Psi_{n}(x_{1},...,x_{n})$ is unique.

From the commutation relation 
\begin{eqnarray}
B_{1}(x)B_{1}(w) & = & \omega(w/x)[B_{1}(w)B_{1}(x)-\frac{1}{y(w/x)}B_{2}(w)A_{1}(x)]\nonumber \\
 &  & +\frac{1}{y(x/w)}B_{2}(x)A_{1}(w)\label{eq4.19}
\end{eqnarray}
we can see that (\ref{eq4.19}) will be normally ordered if it satisfies
the following swap condition 
\begin{equation}
\Psi_{2}(x_{2},x_{1})=\omega(x_{1}/x_{2})\Psi_{2}(x_{1},x_{2})\label{eq4.20}
\end{equation}
This condition fixes $\Gamma(x_{1},x_{2})$ in Eq.(\ref{eq4.18})
and the eigenstate of $\tau(x)$ in the sector $M=2$ has the form
\begin{equation}
\Psi_{2}(x_{1},x_{2})=B_{1}(x_{1})B_{1}(x_{2})\left|0\right\rangle -\frac{1}{y(x_{1}/x_{2})}B_{2}(x_{1})A_{1}(x_{2})\left|0\right\rangle .\label{eq4.21}
\end{equation}
The action of transfer matrix on the states of the form (\ref{eq4.21})
is more laborious. In addition to (\ref{eq4.13a}-\ref{eq4.13}) and
(\ref{eq4.19}) we need appeal to (\ref{eq4.1}) to derive more eight
commutation relations 
\begin{eqnarray}
A_{1}(x)B_{2}(w) & = & \frac{r_{1}(w/x)}{r_{3}(w/x)}B_{2}(w)A_{1}(x)-\frac{r_{7}(w/x)}{r_{3}(w/x)}B_{2}(x)A_{1}(w)\nonumber \\
 &  & +\frac{r_{6}(w/x)}{r_{3}(w/x)}B_{1}(x)B_{1}(w)\label{eq4.22a}\\
A_{2}(x)B_{2}(w) & = & z(x/w)z(w/x)B_{2}(w)A_{2}(x)\nonumber \\
 &  & +\frac{s_{5}(x/w)}{r_{2}(x/w)}[B_{1}(x)B_{3}(w)+B_{3}(x)B_{1}(w)+\frac{s_{5}(x/w)}{r_{2}(x/w)}B_{2}(x)A_{2}(w)]\nonumber \\
\\
A_{3}(x)B_{2}(w) & = & \frac{r_{1}(x/w)}{r_{3}(x/w)}B_{2}(w)A_{3}(x)-\frac{s_{7}(x/w)}{r_{3}(x/w)}B_{2}(x)A_{3}(w)\nonumber \\
 &  & +\frac{1}{y(x/w)}B_{3}(x)B_{3}(w)\\
C_{1}(x)B_{1}(w) & = & -B_{1}(w)C_{1}(x)+\frac{s_{5}(x/w)}{r_{2}(x/w)}[A_{1}(w)A_{2}(x)-A_{1}(x)A_{2}(w)]\\
C_{3}(x)B_{1}(w) & = & -\frac{r_{4}(x/w)}{r_{3}(x/w)}B_{1}(w)C_{3}(x)-\frac{r_{7}(x/w)}{r_{3}(x/w)}B_{1}(x)C_{3}(w)\nonumber \\
 &  & +\frac{1}{y(x/w)}[A_{1}(w)A_{3}(x)-A_{2}(x)A_{2}(w)]+\frac{r_{6}(x/w)}{r_{3}(x/w)}B_{2}(w)C_{2}(x)\nonumber \\
\\
B_{1}(x)B_{2}(w) & = & \frac{1}{z(x/w)}B_{2}(w)B_{1}(x)+\frac{s_{5}(x/w)}{r_{1}(x/w)}B_{1}(w)B_{2}(x)\\
B_{1}(x)B_{3}(w) & = & -B_{3}(w)B_{1}(x)-\frac{s_{5}(x/w)}{r_{2}(x/w)}B_{2}(w)A_{2}(x)+\frac{r_{5}(x/w)}{r_{2}(x/w)}B_{2}(x)A_{2}(w)\nonumber \\
\\
B_{2}(x)B_{1}(w) & = & \frac{1}{z(x/w)}B_{1}(w)B_{2}(x)+\frac{r_{5}(x/w)}{r_{1}(x/w)}B_{2}(w)B_{1}(x)\label{eq4.22}
\end{eqnarray}
Here we obser$w$e that in this approach the final action of $\tau(x)$
on normally ordered states must be normal ordered. This implies in
an increasing use of commutation relations needed for the diagonalization
of $\tau(x)$. For example, the action of the operator $A_{1}(x)$
on $\Psi_{2}(x_{1},x_{2})$ has the form 
\begin{eqnarray}
A_{1}(x)\Psi_{2}(x_{1},x_{2}) & = & z(x_{10})z(x_{20})r_{1}^{L}(x)\ \Psi_{2}(x_{1},x_{2})\nonumber \\
 &  & -\frac{r_{5}(x_{10})}{r_{2}(x_{10})}z(x_{21})r_{1}^{L}(x_{1})\ B_{1}(x)B_{1}(x_{2})\left|0\right\rangle \nonumber \\
 &  & -\frac{r_{5}(x_{20})}{r_{2}(x_{20})}\frac{z(x_{12})}{\omega(x_{12})}r_{1}^{L}(x_{2})\ B_{1}(x)B_{1}(x_{1})\left|0\right\rangle \nonumber \\
 &  & +\left(\frac{z(x_{10})}{\omega(x_{10})}\frac{r_{5}(x_{20})}{r_{2}(x_{20})}\frac{1}{y(x_{01})}+\frac{r_{7}(x_{10})}{r_{3}(x_{10})}\frac{1}{y(x_{12})}\right)\nonumber \\
 &  & \times r_{1}^{L}(x_{1})r_{1}^{L}(x_{2})\ B_{2}(x)\left|0\right\rangle \label{eq4.23}
\end{eqnarray}
where $x_{ab}=x_{a}/x_{b}$, $a\neq b=0,1,2$, with $x_{0}=x$. Here
we have used the following identities satisfied by the matrix elements
of this $19$-vertex model: 
\begin{eqnarray}
\frac{z(x_{ab})}{\omega(x_{ab})}\frac{r_{5}(x_{cb})}{x_{2}(x_{cb})}+\frac{r_{6}(x_{ab})}{r_{3}(x_{ab})}\frac{1}{y(x_{ac})} & = & \frac{r_{5}(x_{ab})}{r_{2}(x_{ab})}\frac{r_{5}(x_{ca})}{r_{2}(x_{ca})}+\frac{z(x_{ac})}{\omega(x_{ac})}\frac{r_{5}(x_{cb})}{r_{2}(x_{cb})},\nonumber \\
z(x_{ab})\frac{r_{5}(x_{cb})}{r_{2}(x_{cb})}\frac{1}{y(x_{ab})}+\frac{r1(x_{ab})}{r_{3}(x_{ab})}\frac{1}{y(x_{ac})} & = & z(x_{ab})z(x_{cb})\frac{1}{y(x_{ac})},\nonumber \\
\omega(x_{ab})\omega(x_{ba}) & = & 1,\qquad\qquad(a\neq b\neq c)\label{eq4.24}
\end{eqnarray}
Similarly, for the operator $A_{2}(x)$ we have 
\begin{eqnarray}
 &  & A_{2}(x)\Psi_{2}(x_{1},x_{2})=\quad\frac{z(x_{01})}{\omega(x_{01})}\frac{z(x_{02})}{\omega(x_{02})}r_{2}^{L}(x)\ \Psi_{2}(x_{1},x_{2})\nonumber \\
 &  & -\frac{s_{5}(x_{02})}{r_{2}(x_{02})}z(x_{21})r_{2}^{L}(x_{2})\ B_{1}(x)B_{1}(x_{1})\left|0\right\rangle \nonumber \\
 &  & -\frac{s_{5}(x_{01})}{r_{2}(x_{01})}\frac{z(x_{12})}{\omega(x_{12})}r_{2}^{L}(x_{1})\ B_{1}(x)B_{1}(x_{2})\left|0\right\rangle \nonumber \\
 &  & +z(x_{21})\frac{1}{y(x_{01})}r_{1}^{L}(x_{1})\ B_{3}(x)B_{1}(x_{2})\left|0\right\rangle \nonumber \\
 &  & +\frac{z(x_{12})}{\omega(x_{12})}\frac{1}{y(x_{02})}r_{1}^{L}(x_{2})\ B_{3}(x)B_{1}(x_{1})\left|0\right\rangle \nonumber \\
 &  & +\frac{s_{5}(x_{01})}{r_{2}(x_{01})}\left(\frac{s_{5}(x_{21})}{r_{2}(x_{21})}\frac{1}{y(x_{01})}+\frac{z(x_{01})}{\omega(x_{01})}\frac{1}{y(x_{02})}-\frac{s_{5}(x_{01})}{r_{2}(x_{01})}\frac{1}{y(x_{12})}\right)\nonumber \\
 &  & \times r_{1}^{L}(x_{2})r_{2}^{L}(x_{1})\ B_{2}(x)\left|0\right\rangle \nonumber \\
 &  & +\frac{1}{y(x_{01})}\left(z(x_{01})\frac{s_{5}(x_{02})}{r_{2}(x_{02})}-\frac{s_{5}(x_{01})}{r_{2}(x_{01})}\frac{s_{5}(x_{02})}{r_{2}(x_{02})}\right)r_{1}^{L}(x_{1})r_{2}^{L}(x_{2})\ B_{2}(x)\left|0\right\rangle \label{eq4.25}
\end{eqnarray}
In this case we have $x$sed more two identities: 
\begin{eqnarray}
\frac{z(x_{ab})}{\omega(x_{ab})}\frac{1}{y(x_{ac})}+\frac{s_{5}(x_{bc})}{r_{2}(x_{bc})}\frac{1}{y(x_{ab})} & = & \frac{s_{5}(x_{ab})}{r_{2}(x_{ab})}\frac{1}{y(x_{bc})}+\frac{z(x_{bc})}{\omega(x_{bc})}\frac{1}{y(x_{ac})}\nonumber \\
z(x_{cb})\frac{s_{5}(x_{ac})}{r_{2}(x_{ac})}+\frac{s_{5}(x_{ab})}{r_{2}(x_{ab})}\frac{s_{5}(x_{bc})}{r_{2}(x_{bc})} & = & \frac{z(x_{ab})s_{5}(x_{ac})}{r_{2}(x_{ac})}\nonumber \\
a & \neq & b\neq c\label{eq4.26}
\end{eqnarray}
Finally, for $A_{3}(x)$ we get 
\begin{eqnarray}
 &  & \left.A_{3}(x)\Psi_{2}(x_{1},x_{2})=\frac{r_{2}(x_{01})}{r_{3}(x_{01})}\frac{r_{2}(x_{02})}{r_{3}(x_{02})}r_{3}^{L}(x)\ \Psi_{2}(x_{1},x_{2})\right.\nonumber \\
 &  & \left.-\omega(x_{12})\frac{1}{y(x_{01})}r_{2}^{L}(x_{1})\ B_{3}(x)B_{1}(x_{2})\left|0\right\rangle -z(x_{21})\frac{1}{y(x_{02})}r_{2}^{L}(x_{2})\ B_{3}(x)B_{1}(x_{1})\left|0\right\rangle \right.\nonumber \\
 &  & +\left(\frac{s_{7}(x_{01})}{r_{3}(x_{01})}\frac{1}{y(x_{12})}-\frac{s_{5}(x_{01})}{r_{3}(x_{01})}\frac{1}{y(x_{02})}\right)r_{2}^{L}(x_{1})r_{2}^{L}(x_{2})\ B_{2}(x)\left|0\right\rangle \label{eq4.27}
\end{eqnarray}
Here we also have $x$sed the identities (\ref{eq4.24}) and (\ref{eq4.26}).

From these relations one can see that all unwanted terms of $\tau(x)\Psi_{2}(x_{1},x_{2})$
vanish. It means that $\Psi_{2}(x_{1},x_{2})$ is an eigenstate of
the transfer matrix $\tau(x)$ with eigenvalue

\begin{equation}
\Lambda_{2}(x,x_{1},x_{2})=z(x_{10})z(x_{20})r_{1}^{L}(x)-\frac{z(x_{01})}{\omega(x_{01})}\frac{z(x_{02})}{\omega(x_{02})}r_{2}^{L}(x)+\frac{r_{2}(x_{01})}{r_{3}(x_{01})}\frac{r_{2}(x_{02})}{r_{3}(x_{02})}r_{3}^{L}(x)\label{eq4.28}
\end{equation}
provided the rapidities $x_{1}$ and $x_{2}$ satisfy the {\small{}BA}
equations 
\begin{equation}
\left(z(x_{a})\right)^{L}=-\frac{z(x_{ab})}{z(x_{ba})}\omega(x_{ba})\ ,\quad a\neq b=1,2.\label{eq4.29}
\end{equation}

\subsection{General Sector}

The generalization of the above results to sectors with more than
two particles proceeds through the factorization properties of the
higher order phase shifts discussed in the previous section. Therefore,
at this point we shall present the general result: In a generic sector
$M=n$ , we have $n-1$ swap conditions 
\begin{equation}
\Psi_{n}(x_{1},\cdots,x_{i-1},x_{i+1},x_{i},\cdots,x_{n})=\omega(x_{i}-x_{i+1})\Psi_{n}(x_{1},\cdots,x_{i-1},x_{i},x_{i+1},\cdots,x_{n})\label{eq4.30}
\end{equation}
which yield the $n-1$ operator-valued functions $\Gamma_{i}(x_{1},\cdots,x_{n})$
. The corresponding normal ordered state $\Psi_{n}(x_{1},\cdots,x_{n})$
can be written with aid of a recurrence formula \cite{TA}: 
\begin{equation}
\Psi_{n}(x_{1},...,x_{n})=\Phi_{n}(x_{1},...,x_{n})\left|0\right\rangle \label{eq4.31}
\end{equation}
where 
\begin{eqnarray}
 &  & \left.\Phi_{n}(x_{1},...,x_{n})=B_{1}(x_{1})\Phi_{n-1}(x_{2},...,x_{n})\right.\nonumber \\
 &  & \left.-B_{2}(x_{1})\sum_{j=2}^{n}\frac{1}{y(x_{1}/x_{j})}\prod_{k=2,k\neq j}^{n}{\cal Z}(x_{k}/x_{j})\Phi_{n-2}(x_{2},...,\stackrel{\wedge}{x}_{j},...,x_{n})A_{1}(x_{j})\right.\label{eq4.32}
\end{eqnarray}
with the initial condition $\Phi_{0}=1,\quad\Phi_{1}(x)=B_{1}(x)$.

The scalar function ${\cal Z}(x_{k}-x_{j})$ is defined by 
\begin{equation}
{\cal Z}(x_{k}/x_{j})=\left\{ \begin{array}{c}
z(x_{k}/x_{j})\qquad\qquad\quad\quad{\rm if}\quad k>j\\
z(x_{k}/x_{j})\omega(x_{j}-/x)\quad\ {\rm if}\quad k<j
\end{array}\right.\label{eq4.33}
\end{equation}

The action of the operators $A_{i}(x),i=1,2,3$ on the operators $\Phi_{n}$
have the following normal ordered form

\begin{eqnarray}
 &  & \left.A_{1}(x)\Phi_{n}(x_{1},...,x_{n})=\prod_{k=1}^{n}z(x_{k}/x)\Phi_{n}(x_{1},...,x_{n})A_{1}(x)\right.\nonumber \\
 &  & \left.-B_{1}(x)\sum_{j=1}^{n}\frac{x_{5}(x_{j}/x)}{x_{2}(x_{j}/x)}\prod_{k=1,k\neq j}^{n}{\cal Z}(x_{k}/x_{j})\Phi_{n-1}(x_{1},...,\stackrel{\wedge}{x}_{j},...,x_{n})A_{1}(x_{j})\right.\nonumber \\
 &  & \left.+B_{2}(x)\sum_{j=2}^{n}\sum_{l=1}^{j-1}G_{jl}(x,x_{l},x_{j})\prod_{k=1,k\neq j,l}^{n}{\cal Z}(x_{k}/x_{l}){\cal Z}(x_{k}/x_{j})\right.\nonumber \\
 &  & \left.\times\Phi_{n-2}(x_{1},...,\stackrel{\wedge}{x}_{l},...,\stackrel{\wedge}{x}_{j},...,x_{n})A_{1}(x_{l})A_{1}(x_{j})\right.\label{eq4.34}
\end{eqnarray}
where $G_{jl}(x,x_{l},x_{j})$ are scalar functions defined by 
\begin{equation}
G_{jl}(x,x_{l},x_{j})=\frac{r_{7}(x_{l}/x)}{r_{3}(x_{l}/x)}\frac{1}{y(x_{l}/x)}+\frac{z(x_{l}/x)}{\omega(x_{l}/x)}\frac{r_{5}(x_{j}/x)}{r_{2}(x_{j}/x)}\frac{1}{y(x/x_{l})}\label{eq4.37}
\end{equation}
For the action of $A_{3}(x)$ we have a similar expression 
\begin{eqnarray}
 &  & \left.A_{3}(x)\Phi_{n}(x_{1},...,x_{n})=\prod_{k=1}^{n}\frac{r_{2}(x/x)}{r_{3}(x/x_{k})}\Phi_{n}(x_{1},...,x_{n})A_{3}(x)\right.\nonumber \\
 &  & \left.+(-1)^{n}B_{3}(x)\sum_{j=1}^{n}\frac{1}{y(x/x_{j})}\prod_{k=1,k\neq j}^{n}{\cal Z}(x_{j}/x_{k})\Phi_{n-1}(x_{1},...,\stackrel{\wedge}{x}_{j},...,x_{n})A_{2}(x_{j})\right.\nonumber \\
 &  & \left.+B_{2}(x)\sum_{j=2}^{n}\sum_{l=1}^{j-1}H_{jl}(x,x_{l},x_{j})\prod_{k=1,k\neq j,l}^{n}{\cal Z}(x_{j}/x_{k}){\cal Z}(x_{l}/x_{k})\right.\nonumber \\
 &  & \times\Phi_{n-2}(x_{1},...,\stackrel{\wedge}{x}_{l},...,\stackrel{\wedge}{x}_{j},...,x_{n})A_{2}(x_{l})A_{2}(x_{j})\label{eq4.36}
\end{eqnarray}
where the scalar f$x$nctions $H_{jl}(x,x_{l},x_{j})$ are given by
\begin{equation}
H_{jl}(x,x_{l},x_{j})=\frac{s_{7}(x/x_{l})}{r_{3}(x/x)}\frac{1}{y(x_{l}/x)}-\frac{s_{5}(x/x_{l})}{r_{3}(x/x_{l})}\frac{1}{y(x/xj)}\label{eq4.50}
\end{equation}
The action of the operator $A_{2}(x)$ is more cumbersome 
\begin{eqnarray}
 &  & \left.A_{2}(x)\Phi_{n}(x_{1},...,x_{n})=(-1)^{n}\prod_{k=1}^{n}\frac{z(x/x)}{\omega(x/x)}\Phi_{n}(x_{1},...,x_{n})A_{2}(x)\right.\nonumber \\
 &  & \left.+(-1)^{n}B_{1}(x)\sum_{j=1}^{n}\frac{s_{5}(x/x_{j})}{r_{2}(x/x_{j})}\prod_{k=1,k\neq j}^{n}{\cal Z}(x_{j}/x_{k})\Phi_{n-1}(x_{1},...,\stackrel{\wedge}{x}_{j},...,x_{n})A_{2}(x_{j})\right.\nonumber \\
 &  & \left.+B_{3}(x)\sum_{j=1}^{n}\frac{1}{y(x/x_{j})}\prod_{k=1,k\neq j}^{n}{\cal Z}(x_{k}/x_{j})\Phi_{n-1}(x_{1},...,\stackrel{\wedge}{x}_{j},...,x_{n})A_{1}(x_{j})\right.\nonumber \\
 &  & +\varepsilon^{n}B_{2}(x)\left\{ \sum_{j=2}^{n}\sum_{l=1}^{j-1}Y_{jl}(x,x_{l},x_{j})\prod_{k=1,k\neq j,l}^{n}{\cal Z}(x_{k}/x){\cal Z}(x_{j}/x)\right.\times\nonumber \\
 &  & \left.\Phi_{n-2}(x_{1},...,\stackrel{\wedge}{x}_{l},...,\stackrel{\wedge}{x}_{j},...,x_{n})A_{1}(x_{l})A_{2}(x_{j})\right.+\nonumber \\
 &  & \left.\sum_{j=2}^{n}\sum_{l=1}^{j-1}F_{jl}(x,x_{l},x_{j})\prod_{k=1,k\neq j,l}^{n}{\cal Z}(x_{l}/x_{k}){\cal Z}(x_{k}/x_{j})\right.\times\nonumber \\
 &  & \left.\Phi_{n-2}(x_{1},...,\stackrel{\wedge}{x}_{l},...,\stackrel{\wedge}{x}_{j},...,x_{n})A_{1}(x_{j})A_{2}(x_{l})\right\} \label{eq4.35}
\end{eqnarray}
where we have more two scalar functions

\begin{eqnarray}
F_{jl}(x,x_{l},x_{j}) & = & \frac{s_{5}(x/x_{l})}{r_{2}(x-x_{l})}\left\{ \frac{s_{5}(x_{l}/x_{j})}{r_{2}(x_{l}/x_{j})}\frac{1}{y(x/x_{l})}+\frac{z(x/x_{l})}{\omega(x/x_{l})}\frac{1}{y(x/x_{j})}\right.\nonumber \\
 &  & \left.-\frac{s_{5}(x/x_{l})}{r_{2}(x/x_{l})}\frac{1}{y(x_{l}/x_{j})}\right\} \label{eq4.38}
\end{eqnarray}
\begin{equation}
Y_{jl}(x,x_{l},x_{j})=\frac{1}{y(x/x_{l})}\left\{ z(x/x_{l})\frac{s_{5}(x/x_{j})}{r_{2}(x/x_{j})}-\frac{s_{5}(x/x)}{r_{2}(x/x_{l})}\frac{s_{5}(x_{l}/x_{j})}{r_{2}(x_{l}/x_{j})}\right\} \label{eq4.39}
\end{equation}

From these relations immediately follows that $\Psi_{n}(x_{1},...,x_{n})$
are the eigenstates of $\tau(x)$ with eigenvalues 
\begin{equation}
\varLambda_{M}(x)=r_{1}(x)^{L}\prod_{a=1}^{n}z(x_{a}/x)-(-1)^{n}r_{2}(x)^{L}\prod_{a=1}^{n}\frac{z(x/x_{a})}{\omega(x/x_{a})}+r_{3}(x)^{L}\prod_{a=1}^{n}\frac{r_{2}(x/x_{a})}{r_{3}(x/x_{a})}\label{eq4.51}
\end{equation}
provided their rapidities $x_{i},i=1,...,M$ \ satisfy the {\small{}BA}
equations 
\begin{equation}
\left(z(x_{a})\right)^{L}=(-1)^{n+1}\prod_{b\neq a=1}^{n}\frac{z(x_{a}/x_{b})}{z(x_{b}/x_{a})}\omega(x_{b}-x_{a}),\quad a=1,2,...,n\label{eq4.52}
\end{equation}
To conclude this section we remark that equations (\ref{eq4.51})
and (\ref{eq4.52}) reproduce the known results in the literature
for the graded nineteen vertex models

\section{Numerical analysis}

For a small-length chain, the results above can be checked numerically.
As an example, let us consider a chain with three sites, that is,
let us assume that $L=2$. In this case, both the monodromy as the
transfer matrix defined at () can be explicitly constructed without
difficult:

The monodromy becomes an operator with values in $\mathrm{End}\left(V_{0}\otimes V_{1}\otimes V_{2}\right)$
and, therefore, consists in an 27-by-27 matrix. Using the graded permutation
operators $P_{01}^{g}=P^{g}\otimes I_{3}$ and $P_{12}^{g}=I_{3}\otimes P^{g}$
we have $\mathcal{R}_{01}=R\otimes I_{3}$ and $\mathcal{R}_{02}=P_{12}^{g}\mathcal{R}_{01}P_{12}^{g}$.
Therefore for two quantum spaces the monodromy (25) is reduced to
\begin{equation}
T_{0}=\mathcal{R}_{02}\mathcal{R}_{01}
\end{equation}

Hence, the graded transfer matrix (\ref{eq4.6}) consists in a $9$-by-9
matrix. 

In the framework of the \noun{aba}, on the other hand, we usually
divide the spectrum of the transfer matrix into sectors, according
to the\emph{ magnon number} $M$ associated with the possible chain
configurations. (We say that a spin pointing to up $\left(1\right)$,
to center $\left(2\right)$, or to down $\left(3\right)$ has a magnon
number equal to 0, 1 or 2, respectively, and that the total magnon
number of the chain is given by the sum of the magnon numbers associated
with all its sites.) In this way, the reference state corresponds
to the sector $M=0$, which physically corresponds to a configuration
in which all spins point up, while the $n$-particle states correspond
to the configurations in which $M=n$, that is, they are physically
formed by any combination of $k$ spins pointing down and $l$ spins
pointing to the center, in such a way that $2k+l=n$. Therefore, for
a chain of length $N=2$, we have in total $5$ sectors, corresponding
to the values of $M$ ranging from $0$ to 4. 

The eigenvalues($\lambda_{i}$ and the eigenvectors ($v_{i})$ of
the transfer matrix can be numerically calculated as we give numerical
values for the parameters $x$, $q$,. In this example, we shall consider
the (randomly generated) values, $x=1.2970895172$ and $q=0.3438435138$. 

Here we notice that for two sites there are nine configurations of
spin$-1$, given by 
\begin{equation}
c_{i,j}=e_{i}\otimes e_{j}\qquad i,j=1,2,3
\end{equation}
 where 
\begin{align}
e_{1}= & \left(\begin{array}{c}
1\\
0\\
0
\end{array}\right), & e_{2}= & \left(\begin{array}{c}
0\\
1\\
0
\end{array}\right), & e_{3}= & \left(\begin{array}{c}
0\\
0\\
1
\end{array}\right)
\end{align}
For this example we find the following eigenvectors their eigenvalues
\begin{alignat}{3}
v_{1}= & c_{1,1},\qquad & \lambda_{1}= & 0.719147295\qquad & (n=0)\nonumber \\
v_{2}= & \frac{1}{\sqrt{2}}\left(c_{2,1}-c_{1,2}\right),\qquad & \lambda_{2}= & -1.1875382707\qquad & (n=1)\nonumber \\
v_{3}= & \frac{1}{\sqrt{2}}\left(c_{2,1}+c_{1,2}\right),\qquad & \lambda_{3}= & 0.80021448548\qquad & (n=1)\nonumber \\
v_{4}= & -\frac{1}{\sqrt{2}}\left(c_{3,1}+c_{1,3}\right),\qquad & \lambda_{4}= & 0.9231082366\qquad & (n=2)\nonumber \\
v_{5}= & -\frac{1}{\sqrt{2}}\left(c_{3,1}-c_{1,3}\right),\qquad & \lambda_{5}= & -1.3365641154\qquad & (n=2)\nonumber \\
v_{6}= & \frac{1}{\sqrt{2}}\left(c_{3,2}-c_{2,3}\right),\qquad & \lambda_{6}= & \lambda_{2}\qquad & (n=3)\nonumber \\
v_{7}= & \frac{1}{\sqrt{2}}\left(c_{3,2}+c_{2,3}\right),\qquad & \lambda_{7}= & \lambda_{3}\qquad & (n=3)\nonumber \\
v_{8}= & c_{2,2},\qquad & \lambda_{8}= & \lambda_{5}\qquad & (n=2)\nonumber \\
v_{9}= & c_{3,3},\qquad & \lambda_{9}= & \lambda_{1}\qquad & (n=4)
\end{alignat}

Now we can look at the results given by the \noun{aba}

The equation (\ref{eq4.11}) for $\text{N=2}$
\begin{equation}
\Lambda_{0}=r_{1}(x)^{2}-r_{2}(x)^{2}+r_{3}(x)^{2}
\end{equation}
Substituiting the numerical values
\begin{alignat}{1}
r_{1}(x)= & -0.84664472310\nonumber \\
r_{2}(x)= & 0.1021523034\nonumber \\
r_{3}(x)= & 0.1159236333
\end{alignat}
to obtain
\begin{equation}
\Lambda_{0}=0.7198147295
\end{equation}

which is equal the value $\lambda_{1}$ the a=eigenvalue obtained
from the transfer matrix in the sector $n=0$.

For the sector $n=1$ we recall the equations (\ref{eq4.16}) and
(\ref{eq4.17}). Now we solve (\ref{eq4.17}) to find $x_{1}$ in
order to find the eigenvalue from $\varLambda_{1}(x,x_{1}$) from
(\ref{eq4.16}). For $N=2$, the numeric solution is
\begin{alignat*}{1}
x_{1}= & -2.9082997350\\
\end{alignat*}
and
\[
\Lambda_{1}(x,x_{1})=0.8002148543
\]
which can be identified with the eigenvalue $\lambda_{3}$ of the
symmetric eigenvector $v_{3}$ of the transfer matrix for the sector
$n=1.$

For the sector $n=2$, we recall the equations (\ref{eq4.28}) and
(\ref{eq4.29}) with $L=2.$

The two equations supplied by (\ref{eq4.29}) can be numerical solved
to find $x_{1}$ and $x_{2}$ and then we can find $\varLambda_{2}(x,x_{1},x_{2})$
from (\ref{eq4.28}). 

The numerical results are 
\[
x_{1}=-2.9082997360
\]
\[
x_{2}=-2.9082997340
\]
and
\[
\Lambda_{2}(x,x_{1},x_{2})=0.9231082388
\]
which is the eigenvalue $\lambda_{4}$ of the symmetric eigenvector$v_{4}$
of the transfer matrix.

we hope these few examples should be suffice to pave the way for $L\geq3$
. 

We remark however that only 5 of the 9 eigenvalues of the transfer
matrix are actually distinct, which is due to the symmetry of the
system regarding inversion of the spins.

In order to compute the eigenvalues of the transfer matrix in the
framework of the \noun{aba}, we need to solve the \noun{bae}, since
the eigenvalues given by (\ref{eq4.28}) depend implicitly on the
rapidities -- \emph{i.e.}, on the solutions of the\noun{ bae}. Here
we remark as well that for $L=2$ is not necessary to go up to $n=4$,
as we could expect from (76). The solutions for $n=\{3,4\}$ provide
the same eigenvalues as that obtained from the cases $n=\{2,1\}$,
respectively, which is due to the above mentioned symmetry of the
system regarding the inversion of the spins. This is very welcome,
since the \noun{bae} are very difficult to solve, even numerically.
In fact, the \noun{bae} are highly ill-conditioned: their roots are
very close to each other, which requires a high accuracy in the computations;
there are solutions which are not physical (for instance, in the present
case when case some root equals $0$, $\pm1$, $\pm1/q^{2}$, or when
two or more roots are equal to each other etc.) and they should be
discarded. Different solutions may lead to the same eigenvalue, for
example those solutions differing only by a permutation of the Bethe
roots, or, sometimes, roots differing only by a complex conjugation.
For more details about the complexity \noun{bae}, (see \cite{VLS}).

\section{Conclusion}

In this work we derived the periodic algebraic \textsc{ba} for the
supersymetric nineteen vertex model constructed from a three-dimensional
free boson representation $V$ of the twisted quantum affine Lie superalgebra
$U_{q}[\mathrm{osp}(2|2)^{(2)}]\simeq U_{q}[C(2)^{(2)}]$. Explicit
results and a numerical analysis were also presented.

\end{document}